\documentclass[twocolumn,showpacs,preprintnumbers,amsmath,amssymb,prl]{revtex4}

\usepackage{graphicx}
\usepackage{dcolumn}
\usepackage{bm}

\begin{document}

\title{Absence of Dipole Glass Transition for Randomly Dilute Classical
Ising Dipoles}

\author{Joseph Snider$^{\dagger}$ and Clare C. Yu}
\affiliation{Department of Physics, University of California,
    Irvine, California 92697-4575}
\date{\today}

\begin{abstract}
Dilute dipolar systems in three dimensions are expected to undergo a
spin glass transition as the temperature decreases. Contrary to
this, we find from Wang-Landau Monte Carlo simulations that at low
concentrations $x$, dipoles randomly placed on a cubic lattice with
dipolar interactions do not undergo a phase transition. We find that
in the thermodynamic limit the ``glass'' transition temperature
$T_g$ goes to zero as $1/\sqrt{N}$ where $N$ is the number of
dipoles. The entropy per particle at low temperatures is larger for
lower concentrations ($x=4.5\%$) than for higher concentrations
($x=20\%$).
\end{abstract}

\maketitle

Disordered insulating materials often have randomly placed electric
or magnetic dipoles that have long range dipolar interactions.
Examples include impurities in alkali halides that can be used for
paraelectric cooling \cite{Kanzig1964,Narayanamurti1970}, diluted
ferroelectric materials \cite{Vugmeister90}, disordered magnetic
materials, and frozen ferrofluids \cite{Luo91}. These systems are
typically modeled as spin glasses that have simpler interactions and
yet are believed to capture the essential physics of interacting
dipoles. Based on theoretical studies of spin glasses with long
range interactions \cite{Stephen1981,Xu1991,Yu1992,Katzgraber2003},
one would expect dilute Ising dipolar systems to undergo a spin
glass-like transition as the temperature decreases. However, in this
paper we find the surprising result that, unlike the three
dimensional Ising spin glass with $1/r^{3}$ interactions
\cite{Katzgraber2003,Bray1986}, dilute Ising dipolar systems 
do not undergo a phase
transition as the temperature decreases. This may explain the lack
of experimental evidence for such a transition in very dilute
systems.

An example of dipoles is two level systems (TLS) that dominate the
physics of glasses at low temperatures \cite{esquinazi}. TLS often
have randomly oriented electric dipole moments that interact through
an elastic strain field with a long range interaction that is a
stress tensor generalization of the vector dipolar
interaction\cite{Yu1989}. While there have been experimental hints
of a spin glass transition among TLS in glasses at low temperatures
\cite{Strehlow98}, there has been no definitive experimental proof
that such a transition occurs. Since the estimated concentration of
TLS is low (100 ppm), our result may explain the absence of a
transition even though TLS dipoles are randomly
oriented and may not be Ising.

Another example is the insulator LiHo$_x$Y$_{1-x}$F$_4$
\cite{Reich1990} in which the holmium ions have Ising magnetic
dipole moments that lie along the z-axis due to crystal field
effects \cite{Hansen1975}. For very dilute systems ($x=4.5\%$)
LiHo$_x$Y$_{1-x}$F$_4$ shows no sign of a transition
\cite{Ghosh2002}. The lack of low temperature freezing in
LiHo$_x$Y$_{1-x}$F$_4$ has been attributed to dominant quantum
mechanical effects in the so-called spin liquid or antiglass phase
\cite{Ghosh2002,Ghosh2003}. However, a theoretical investigation of
whether or not classical interacting dipoles undergo a spin glass
phase transition at low concentrations has been lacking. Several
previous studies of dipolar interactions between randomly placed
Ising dipoles have focussed on the ferromagnetic transition that
occurs at higher dipole concentrations
\cite{Stephen1981,Zhang1995,Ayton1997}. Monte Carlo simulations have
looked at intermediate concentrations with $x\geq 25\%$ where there
is a spin glass transition \cite{Yu1992,Jensen1989}. Xu {\it et al.}
\cite{Xu1991} used mean field theory and found, depending on the
lattice structure, ferromagnetic or antiferromagnetic transitions at
higher concentrations. They found a spin glass phase at lower spin
concentrations but the properties of this phase were unreliable
because they had a replica symmetric solution. In short, there have
been no definitive theoretical studies of the very dilute classical
cases. In this paper we present the results of Wang-Landau Monte
Carlo simulations on classical dilute Ising dipolar systems in three
dimensions. We find that there is no phase transition for low
concentrations in qualitative agreement with experiment.

In spin glasses the distribution $P(q,T)$ of the overlap order
parameter $q$ changes from being a Gaussian centered at $q=0$ at
high temperatures to a bimodal distribution with peaks at $q=\pm 1$
at low temperatures. At intermediate temperatures it is relatively
flat. We can define a characteristic glass transition temperature
$T_g$ as the temperature where $P(q,T)$ is the flattest. In the
thermodynamic limit we find that for a given dipole concentration
$T_g$ goes to zero as $1/\sqrt{N}$ where $N$ is the number of
dipoles. Also, we examine the entropy and find that for
concentrations less than $20\%$ there is a nonzero entropy per dipole as
$T\rightarrow0$. The entropy and lack of a transition are consistent
with a large number of accessible low energy states and glassy
behavior.

We consider Ising dipoles randomly placed on a simple cubic lattice
at concentrations of $x=4.5\%$, $12\%$, and $20\%$. The interaction
between any two dipoles $\vec{p}_{1}$ and $\vec{p}_{2}$ separated by
a vector $\vec{r}_{12}$ is given by the Hamiltonian:
\begin{equation}
   \label{eq:Hamiltonian}
   H(\vec{p}_{1}, \vec{p}_{2}) =
      \frac{\vec{p}_{1} \cdot \vec{p}_{2} -
         3(\hat{r}_{12}\cdot\vec{p}_{1})(\hat{r}_{12}\cdot\vec{p}_{2}) }
      {r_{12}^{3}}.
\end{equation}
In addition to the energy units set by $H$, the units are set by
$\vec{p}_{i}=\pm\hat{z}$, the lattice constant $a=1$, and
Boltzmann's constant $k_B=1$. These will be referred to as MC units
where appropriate. We use the Ewald summation technique to handle
the long range nature of the dipole interactions \cite{DeLeeuw1980}.
For $x=4.5\%$, the lattices had $L^3$ sites with $L=6$, $8$, $10$,
and $12$, and for $x=12\%$ and $20\%$, $L=4$, $6$, and $8$. The
number of dipoles $N$ is the smallest even integer greater than or
equal to $xL^{3}$.

The glassy energy landscape at low concentrations makes it difficult
to equilibrate at low temperatures with the traditional Metropolis
Monte Carlo approach. To overcome this, we have used the Wang-Landau
(WL) Monte Carlo technique \cite{Wang2001} to calculate the density
of states $n(E)$ where $E$ is the energy of the system. Briefly,
this algorithm starts with an initial guess $n(E)=1$ and executes a
weighted random walk on the energy landscape. Single flips of
randomly selected dipoles are then accepted with a probability of
$\min\left[1, n(E_{i})/n(E_{f})\right]$ where $E_{i}$ and $E_{f}$
are the energies before and after the trial flip. If a step is
accepted (rejected), then the density of states is updated by the
rule $n(E_{f(i)}) \rightarrow \gamma n(E_{f(i)})$ where $\gamma>1$
is a scale factor. A histogram of the visited energies $h(E)$ is
recorded. The criterion for a satisfactory estimate of the density
of states is given by the flatness of $h(E)$, \textit{i.e.},
$h(E)>\epsilon\left<h\right>$ for every energy $E$ where
$0<\epsilon<1$ determines the accuracy; typically,
$\epsilon\approx0.95$. Once the flatness condition is satisfied, the
scale factor is set closer to $1$ by the rule
$\gamma\rightarrow\sqrt{\gamma}$, $h(E)$ is reset to zero, and the
algorithm is repeated. In all cases, we ran $20$ iterations with
$\gamma$ starting at $e$ and ending at $1.0000019$, and $n(E)$ was
normalized such that $\sum_{E}n(E)=2^{N}$.

The dipolar interaction is nearly continuous so each energy bin may
contain multiple states.
We choose the bins to be as small as possible while maintaining
reasonable computational times. The bin sizes depend
on concentration and system size. The bins are about $0.01$ in units of
energy per particle for $20\%$ filling and $0.001$ for $4.5\%$ and
$12\%$. The lowest temperature studied ($T=0.05$) must be larger 
than the largest bin (0.02). We try to keep the bins small enough so that 
$n(E_{0})\sim 2$, where $E_{0}$ is the energy of the (degenerate) 
ground state. In all cases except one $n(E_{0})\approx3.5$.
The exception ($8^{3}$ at $20\%$) has $n(E_{0})\approx 10$, so we
discard the low temperature values of this system.

We average over disorder by having different runs correspond to
different quenched placements of dipoles with random initial 
orientations. The
dipoles are fixed in position but not in orientation. There are
about $1000$ runs for each $x$ and $L$. As a check of our Wang-Landau
procedure, we were able to enumerate all the states for $1000$
different configurations for concentrations of $4.5\%$ ($L=6$ and
$8$), $12\%$ ($L=4$) and $20\%$ ($L=4$) and determine the exact
density of states. We found very good agreement with our WL results.

Since we are looking for a spin glass phase, we define a generalized
Edwards-Anderson overlap order parameter $q=\frac{1}{N}\sum_{i}
\vec{p}_{i}^{\;g} \cdot \vec{p}_{i}^{\;s}$, where
$\vec{p}_{i}^{\;s}$ is a dipole in the state of the current system,
and $\vec{p}_{i}^{\;g}$ is a dipole in a low energy state found in a
short, initial simulation \cite{Bhatt1988,Yamaguchi2003}. Then, to
find the distribution $P(q,E)$, $q(E)$ is sampled and stored in a
histogram during the simulation at the smallest scale factor where
the estimate of the density of states is quite good. $P(q,T)$ is
calculated as
\begin{equation}
P(q,T)=C_{T}(1/Z)\sum_{E}n(E) P(q,E)\exp(-E/kT)
\end{equation}
where the sum is over all the energy bins and $Z(T)=\sum_{E}
n(E)\exp(-E/kT)$. $C_{T}$ enforces normalization such that
$\sum_{q}P(q,T)=1$ for every $T$. This method has been seen to give
a reasonable order parameter distribution in the case of a Potts
model \cite{Yamaguchi2003}.

It has often been convenient to find the spin glass transition
temperature using Binder's $g=[3-\langle \left(q^4\rangle/\langle
q^2\rangle^2\right)]/2$, and $\langle q^{m}\rangle = \sum_{q} q^{m}
P(q,T)$ \cite{Binder1981}. Since the ground state estimate is not
the true ground state, we eliminate all runs in which $g<0.8$ at the
lowest temperature. If there is a second order phase transition,
plots of $g$ versus $T$ for different size systems will cross at the
transition temperature \cite{Bhatt1988}. However, we find that these
curves do not cross, so there is no second order spin glass phase
transition.

\begin{figure}[tbp]
   \begin{center}
      \includegraphics[width=3in]{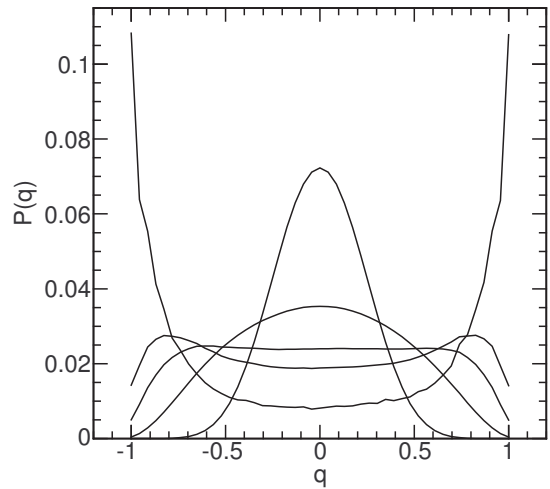}
\caption{$P(q)$ at $x=4.5\%$, $L=10$ (46 dipoles). $T=$5, 1.6, 1.1,
0.9, and 0.5. The lines transition from a Gaussian ($T=5$) to a
bimodal distribution ($T=0.5$).}
      \label{fig:P0WL}
   \end{center}
\end{figure}

\begin{figure}[tbp]
   \begin{center}
      \includegraphics[width=3in]{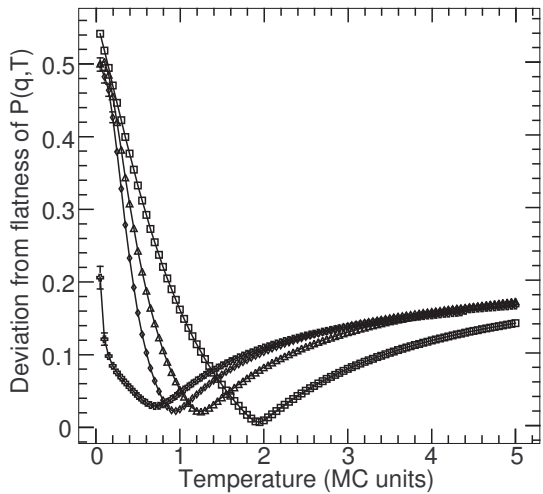}
\caption{Deviation from flatness of $P(q,T)$ for $x=4.5\%$ with
sizes $L=6^{3}$, $8^{3}$, $10^{3}$, and $12^{3}$. The minima moves
left with increasing size and defines a glass transition
temperature. The plots for $x=12\%$ and 20\% are similar.}
      \label{fig:pq_flatness}
   \end{center}
\end{figure}

\begin{figure}[tbp]
   \begin{center}
      \includegraphics[width=3in]{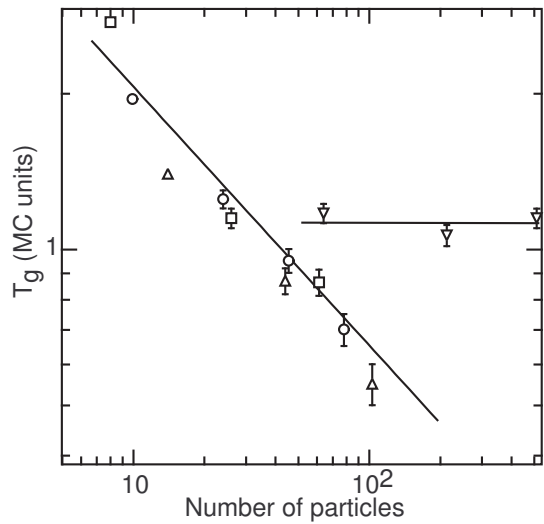}
\caption{Log-log plot of the maximum flatness for $P(q)$ versus the number of 
dipoles at various concentrations. Open
circles are $4.5\%$, squares are $12\%$, triangles
are $20\%$, and upside down triangles are 100\%. 
The left solid line has a slope of $-1/2$ corresponding to
$T_g \sim N^{-1/2}$. The right solid line has a slope of zero.
Fits to $T_g \sim N^{-\alpha}$ yield $\alpha=0.49(1)$ for 4.5\%,
$\alpha=0.6(1)$ for 12\%, $\alpha=0.45(3)$ for 20\%, and
$\alpha=0.02(5)$ for 100\%. The errors in the last digit are in
parentheses.
} 
      \label{fig:pq_max_flatness}
   \end{center}
\end{figure}

To investigate this further, we can look at how $P(q,T)$ changes
with temperature. For a system undergoing
a phase transition, we expect $P(q,T)$ to change from being a
Gaussian centered at $q=0$ at high temperatures to a bimodal
distribution with peaks at $q=\pm 1$ at low temperatures. A
typical example is shown in Figure \ref{fig:P0WL} for $x=4.5\%$,
$L=10$. We define a characteristic ``glass'' temperature $T_g$
as the temperature where the distribution $P(q,T)$ is flattest. We
define the deviation $D(T)$ from flatness in terms of the variance
of $P(q,T)$ as
\begin{equation}
   \label{eq:DeviationFromFlatness}
      D(T) = L^{3}\left\langle\left(P(q,T)-
             \left\langle P(q',T)\right\rangle_{q'}\right)^{2}
             \right\rangle_{q}
\end{equation}
where $\left\langle\ldots\right\rangle_{q}$ indicates an average
over all $N+1$ possible values of $q$. $D(T)$ is at a minimum when a
plot of $P(q,T)$ versus $q$ is the flattest, defining $T_{g}$. A
sample $D(T)$ is plotted in figure \ref{fig:pq_flatness}. For a
given dilute concentration, $T_{g}$ is tending to smaller temperatures as
the system size increases which is consistent with $T_{g}\rightarrow
0$ as $L\rightarrow\infty$. 
To find the dependence of $T_g$ on the number $N$ of dipoles for a
fixed concentration, we plot the minimum of $D(T)$ versus $N$ in
figure \ref{fig:pq_max_flatness}. The best fit for dilute cases
reveals that $T_{g}\sim N^{-1/2}$. In contrast, $D(T)$ for the
ordered case ($x=100\%$) yields a nonzero transition temperature
independent of $N$.

The absence of a transition is consistent with
the experimental finding that for very dilute systems ($x=4.5\%$)
LiHo$_x$Y$_{1-x}$F$_4$ shows no sign of a transition
\cite{Ghosh2002}. 
However, the absence of a transition in dilute dipolar systems is unexpected
since 3D Ising spin glasses with $1/r^{3}$ interactions undergo a
phase transition \cite{Katzgraber2003,Bray1986}. $P(q)$ 
for a spin glass and for a dilute dipolar system are different; 
in the
thermodynamic limit as $T\rightarrow 0$, $P(q)$ for a spin glass has
a few sharp peaks corresponding to ground state configurations
separated by high barriers, while $P(q)$ for the dilute dipolar
system is flat, indicating numerous accessible low energy states
separated by insignificant barriers. With very low barriers, states
at both the top and bottom of the barrier contribute low energy
states. The difference in barrier heights may be due to every site
in a model spin glass being occupied so that in a spin glass with
power law interactions nearby spins will tend to have stronger
interactions than distant spins and produce large barrier heights.
In a dilute dipole system nearby sites are empty and so the low
energy configurations are determined by distant dipoles which
interact weakly and produce low barriers.

The presence of many nearly degenerate accessible ground states is
reflected in the finite entropy per dipole near $T=0$. We find that
the low temperature entropy is larger for the lower concentration.
We can calculate the total entropy $\mathcal{S}_{tot}(T)$ directly
from the density of states obtained by our WL Monte Carlo
simulations:
\begin{equation}
   \mathcal{S}_{tot}(T)=\frac{\langle E(T)\rangle}{T}+\log Z(T)
\end{equation}
where $\langle E(T)\rangle$ is the average energy of the system. 
$\mathcal{S}_{tot}$ is an absolute entropy and is not defined relative 
to some reference value. To
compare different system sizes, we consider the entropy per particle
$S_{N}=\mathcal{S}_{tot}/N$ where $N$ is the number of dipoles. The
entropy is very smooth, corresponding to a broad bump in the
specific heat.

\begin{figure}[tbp]
   \begin{center}
      \includegraphics[width=3in]{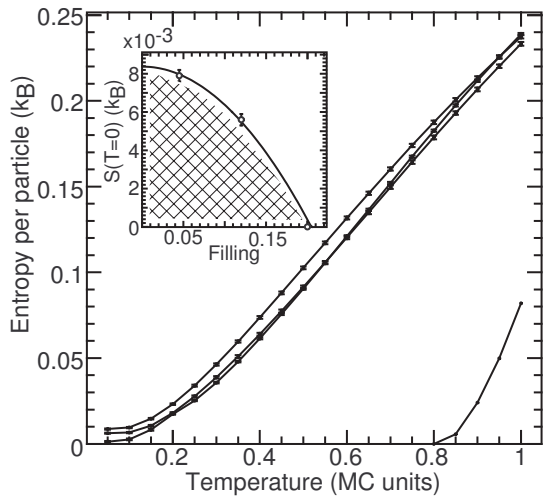}
\caption{Entropy per particle extrapolated to infinite size. From
top to bottom the fillings are $4.5\%$, $12\%$, and $20\%$. The 
curve on the far right is for 100\%. The
error bars shown represent the standard error ($\sim 10^{-3}$)
of the distribution of $S(T)$. Inset: Entropy at $T=0$ phase 
diagram. The hashed area is classically not accessible. The 
solid line is a guide to the eye. Above $20\%$ $S(T=0)$ is zero.}
      \label{fig:WL_Zero_Entropy_Extrapolated}
   \end{center}
\end{figure}

To determine the entropy in the thermodynamic limit, we plot
$S_{N}(T)$ versus $1/N$ at a given temperature $T$. We fit a line to
the data and then extrapolate to $N\rightarrow\infty$. Then, we plot
the extrapolated value versus temperature (see Figure
\ref{fig:WL_Zero_Entropy_Extrapolated}). From Figure
\ref{fig:WL_Zero_Entropy_Extrapolated} it is clear that the 4.5\%
and $12\%$ cases have a nonzero entropy at low temperatures, but the
$20\%$ case is approaching zero. Finally, the extrapolated values
are fit with a power law of the form $A T^{\lambda}+S_o$, where $A$
and $\lambda$ are constants, and $S_o$ is a constant representing
the zero temperature value of the entropy. The fit values are
$A=1.1\pm0.2$, $\lambda=2.7\pm0.1$, and
$S_o=(7.9\pm0.3)\times10^{-3}$ at $4.5\%$, $A=1.2\pm0.3$,
$\lambda=2.9\pm0.2$ and $S_o=(5.6\pm0.3)\times10^{-3}$ at $12\%$,
and $A=0.37\pm0.05$, $\lambda=1.9\pm0.1$ and
$S_o=(-0.5\pm0.5)\times10^{-3}$ at $20\%$. Note, the extrapolation
at $20\%$ gives a negative $S_o$, so it is zero; no actual data
points have negative entropy. A phase diagram of the entropies at
zero temperature is constructed in the inset of Figure
\ref{fig:WL_Zero_Entropy_Extrapolated}. Notice that the low temperature
entropy increases as the concentration decreases. This 
indicates that there are more accessible low energy states in
systems with lower concentrations where the dipoles interact more
weakly. Having a finite value of
$S_o$ implies that the zero temperature entropy $S(0)$ may be
nonzero, but this is not unprecedented for a classical system, e.g.,
noninteracting spins.

We do not think that the finite entropy near $T=0$ is due to
the finite size of the energy bins. To test the effect of the bin
size, we halved the bin size (doubled the number of bins) for
the case of $6^{3}$ at $20\%$ and found an entropy change of about
5\% which is consistent with the error estimates. We also ran 
the largest exact case ($8^{3}$
at $4.5\%$) through the WL algorithm with bins of width
$0.005$, and found a change of $0.9\%$ compared to the
exact result with zero bin width.

To summarize, we 
find the surprising result that at low concentrations ($x\leq 20\%$)
there is no spin glass-like phase transition as the temperature is
lowered. This is consistent with having a large number of nearly
degenerate accessible low energy states. Our result could explain
the lack of experimental evidence for a transition in
LiHo$_x$Y$_{1-x}$F$_4$ for small $x$ and among two level systems in
glasses at low temperatures. Thus,
contrary to widely held notions, materials with dilute
electric or magnetic dipoles cannot necessarily be modeled as spin
glasses with long range interactions. 

We thank Stuart Trugman and Manoranjan Singh for helpful
discussions. This work was supported by DOE grants DE-FG03-00ER45843
and DE-FG02-04ER46107.

$^{\dagger}$Present address: The Salk Institute, La Jolla, CA 92037


\end{document}